\documentclass[twocolumn]{revtex4}

\usepackage{graphics}
\usepackage{amssymb,epsf}

\begin{document}
\title{\Large{\textbf{Fine Structure of Avalanches in the Abelian Sandpile Model}}}
\author{Amir Abdolvand}

\author{Afshin Montakhab\footnote[1]{Corresponding author. Tel.: +98-711-2284609; Fax: +98-711-2284594 \\E-mail address: montakhab@shirazu.ac.ir}}

\affiliation{Physics Department, College of Science, Shiraz University, Shiraz 71454, Iran.}

\begin{abstract}
We study the two-dimensional Abelian Sandpile Model on a square lattice of linear size L. We introduce the notion of avalanche's
fine structure and compare the behavior of avalanches and waves of toppling. We show that according to the degree of complexity in
the fine structure of avalanches, which is a direct consequence of the intricate superposition of the boundaries of successive waves,
avalanches fall into two different categories. We propose scaling ans\"{a}tz for these avalanche types and verify them numerically.
We find that while the first type of avalanches has a simple scaling behavior, the second (complex) type is characterized by an avalanche-size dependent scaling exponent. This provides a framework within which one can understand the failure of a consistent scaling behavior in this model.\\

\begin{flushleft}
\end{flushleft}
\end{abstract}
\maketitle
\section{Introduction}
Bak, Tang, and Wiesenfeld (BTW) introduced the notion of
Self-Organized Criticality (SOC) [1,2] as a possible mechanism for
the generic emergence of spatial and temporal power law
correlations. To elucidate the concept of SOC, they introduced acellular automaton known as sandpile model which is an example of
slowly driven, spatially extended, dissipative dynamical system
[3]. The generality inherent in the basic notions of SOC has led
to its successful application in various problems in physics
as well as biology [4,5,6,7]. The common characteristics of all
these systems is that at the self-organized critical state the
microscopic details of the system are shadowed by the collective
behavior of the individual constituents of the system. Due to the
simplicity of local dynamical rules, and the ease with which they
are implemented on a computer, many different models exhibiting
SOC have been introduced and studied by various authors[8,9].
However, the prototypical sandpile model of SOC and a variant of
it known as the Abelian Sandpile Model model (ASM)[10], has
resisted many clever efforts in fully understanding its dynamical
behavior [11,12,13,14,15]. This is despite the fact that the
analytical tractability of this model, which enables one to
evaluate exactly many of its static properties [10,16,17,18], had
provided hope for a better understanding of its dynamical
properties.  Currently, a complete description of the dynamical
properties of the ASM is still missing [19].

To highlight this important point, we concentrate on the
event-size (avalanche) distribution function. In order to check
the assumption that the characteristic properties of avalanches in
the critical state are described by scale free distribution
functions with cutoffs limited only by the finite size effects,
BTW proposed a simple picture of finite-size scaling (FSS) in
analogy with more traditional critical phenomena [3,20,21].
Although, it is now generally accepted that simple FSS picture
fails in describing the scaling behavior of avalanches in the BTW
model, the reasons suggested for this inconsistency seem to be
very different [22,23,24,25,26]. For example, in a large-scale
simulation, Drossel [26] explained the deviations from pure power
law behavior by dividing avalanches into dissipative and
non-dissipative avalanches. However, such attempts which aim at
relating the deviations from pure power law behavior to the
finite-size (or boundary) effects seem to be problematic. In fact,
it is shown by Ktitarev et.\ al [27], that by reducing such
effects one still observes the aforementioned deviations from the
simple power law behavior.

Due to the complex spatiotemporal behavior of avalanches, it is
reasonable to decompose them into more elementary objects.  As is
shown in Ref.\ [28], due to the Abelian property of the model
which admits an interchangeable order in the relaxation of local
instabilities, one can consider an avalanche as a composition of a
series of (global) instabilities which are referred to as waves of
topplings. The inconsistencies in our understanding of the
dynamical behavior of avalanches reveal themselves more clearly
when we consider that an ensemble of waves behaves simply and
obeys FSS ans\"{a}tz [27]. This is in contrast with the complex
behavior of avalanches. That is, avalanches which are
\textit{presumably} simple composition of waves do not obey FSS.
In view of the aforementioned points, the following main questions
arise: While the time evolution of an avalanche differs only in
the order of the relaxation of local instabilities with that of a
wave, what makes the two events have such different scaling
behaviors? Moreover, beside peculiarities imposed by boundaries of
the system and finite size effects, what possible mechanism,
probably inherent in the dynamical behavior of an avalanche
itself, might be responsible for the observed complexity in the
scaling behavior of avalanches?

In general, one expects to answer these questions by considering
the effect of ``nontrivial composition'' of correlated waves
[29,30]. However, to obtain a clear quantitative picture one needs
to clarify beforehand what exactly ``nontrivial composition''
means, how this nontriviality is related to the complexity in
avalanche dynamics, and last but not least, how this is related to
the failure of simple scaling picture [31]. In the present work,
we analyze the spatiotemporal structure of avalanches in order to
investigate the above issues. The bulk of an avalanche consists of
sites which have toppled as well as their nearest neighbors. This
might lead one to naively believe that the bulk of an avalanche
consists of sites which are unchanged, i.e. recurrent, where only
sites on the boundary of an avalanche change their states
(dynamical variable). However, avalanches can have complex
internal structures. As waves of topplings occur during an
avalanche, the boundaries of these waves could interact with each
other leading to a complex internal (bulk) structure consisting of
both recurrent and non-recurrent states, see for example Fig.
1(b). Using these facts, we classify avalanches into two classes,
simple and complex, and investigate their scaling behavior. We
show that different classes of avalanches have distinctly
different scaling behavior. In particular, while the scaling
behavior of the first type is observed to be independent of
avalanche-size, in the second type an avalanche-size dependent
scaling exponent is found. We therefore argue this to be the main
cause of inconsistent scaling behavior in avalanche statistics of
the ASM.

The present article is organized as follows: In Section II, we
give a brief review of the basic concepts and definitions of the
ASM. In Section III, we analyze the spatial structures of
avalanches and introduce the idea of the avalanche's fine
structure. In Section IV, we compare the behavior of avalanches
and waves of topplings and argue that according to the degree of
complexity in their fine structures, avalanches fall into two
different categories, type $\alpha$ and type $\beta$.  In Section
V, we propose scaling ans\"{a}tz for the two types of avalanches
and verify them numerically. Finally, Section VI is devoted to a
short summary and outlook.
\section{Two-dimensional ASM and the wave picture of evolution}
The two-dimensional ASM is a cellular automaton defined on a
square lattice of linear size L. To every point of the lattice
there corresponds an integer dynamical variable $h_{i}$, which in
the language of sandpiles represents the height of the column of
sand at the $i^{th}$ site. To simulate external drive, the system is
perturbed by increasing the dynamical variable of a randomly chosen site by one,
\begin{equation}
    h_{i} \rightarrow h_{i}+1.
\end{equation}
This can be interpreted as an increase in the local value of
height, energy, pressure, etc. A site is considered unstable if
its dynamical variable exceeds a predefined threshold value
$(h_{i}>h_{c})$. An unstable site then topples, upon which its
dynamical variable is decreased by 4, whilst each of its four
nearest neighbors $(nn)$ receive one unit of energy:
\begin{eqnarray}
h_{i}& \rightarrow & h_{i}-4,\\
h_{nn}& \rightarrow & h_{nn} +1.
\end{eqnarray}
In turn, through the relaxation processes, see Eqs.(3), the
neighboring sites may become unstable themselves, leading to a
series of instabilities the sum of which is referred to as an
avalanche. Since the local dynamical rules are conservative, the
dissipation can take place only at the boundary of the system.
Here we will use open boundary conditions where, if an unstable
site is on the boundary, one or more grains of sand will leave the
system.

The method generally used during the relaxation of an avalanche is
the parallel updating method, where all unstable sites are relaxed
simultaneously during an instant in the relaxation process. Due to
the Abelian nature of the model, the order of the toppling during
an avalanche does not affect the final state [10]. Therefore,
beside the parallel method of updating, it is possible to perform
the relaxation process by a succession of waves. There is a simple
dynamical procedure leading to such a decomposition [28]. In this
method, we relax the seeding site, say $i$, after its first
instability. This may cause further instabilities in the
neighboring sites. We then relax all other unstable sites, except
the seeding site $i$. The set of all toppled sites during this
process forms the first wave of toppling. If, after the
termination of the first wave, the seeding site is still unstable,
we repeat the above procedure obtaining the second, third, etc.\
waves of topplings. This procedure continues until all sites are
stable again. Therefore, one can consider the relaxation process
of an avalanche as a sequence of waves of topplings, all of which
originate from the seeding site. While in the former method waves
overlap in time, by decomposing an avalanche to a sequence of
waves, only one wave propagates at a time. This method of updating
enables us to view the time evolution of the model in an ensemble
of waves.

It can be shown both analytically and numerically [27] that the
scaling property of waves is simple and obeys FSS\@. However,
avalanches, which might naively be considered as a simple sum of
waves, do not have simple scaling behavior and in fact do not obey
FSS\@ [22,23,27]. We believe that the reason for this discrepancy is
found in the fine structure of avalanches which is due to wave
boundary interactions.

In the present work we show that the boundaries of waves making up
a given avalanche could interact with each other. This interaction
of boundaries may lead to a complex spatial structure within an
avalanche bulk structure which substantially changes the dynamical
properties of the avalanche, including its scaling behavior. To
show this, we divide avalanches into two distinct classes based on
the complexity in their (internal) fine structure: simple
(type $\alpha$) avalanches that behave much like waves, and
complex (type $\beta$) avalanches which have distinctly different,
size-dependent scaling behavior.

\section{Fine Structure of an Avalanche}
The area \textit{a} of a relaxation process $R$ (avalanche or
wave) is defined as the number of distinct sites toppled during
that process. In general, this area can be divided into two
different structures which make up the fine structure of an
avalanche. The first structure consists of those sites for which
the corresponding dynamical variables have remained unchanged
before and after the relaxation event. In fact, these sites may
have toppled once or more during the relaxation event. But due to
the balance between the outflow (Eq.\ (2)), and inflow (Eq.\ (3))
of particles through the relaxation of their nearest neighbors,
their states remain unchanged after the termination of the
relaxation process. The second structure consists of sites whose
dynamical variables have changed as a result of a relaxation
event.

Let us denote by \( |h_{i},\ t\rangle,\ i=1,\ldots,N \) the single
site microstates of the system at time \textit{t}, where
\textit{t} denotes the macroscopic time scale of the system. We
denote by \( |h_{i},\ t\rangle_{I} \) and \( |h_{i},\ t\rangle_{F}
\), the initial and final microstate of the $i^{th}$ site before
and after the occurrence of the $t^{th}$ relaxation process. A
Recurrent Macrostate $\mathcal{(RM)}$ of the system consists of
those sites for which the corresponding single site microstates
satisfy the following relation,
\begin{equation}
    i\in \mathcal{(RM)}\ \Leftrightarrow\ |h_{i},\ t\rangle_{F}=|h_{i},\ t\rangle_{I} \wedge i\in R
\end{equation}
We say that the site \textit{i} belongs to the relaxation process
$R$, $i \in R$, if and only if \textit{i} has toppled during that
process. $R$ is the region affected by the process.

In a similar way, one can define the Non-recurrent Macrostate
$\mathcal{(NM)}$ of a relaxation process as a collection of sites
for which the corresponding single site microstates satisfy the
relation,
\begin{equation}
    i\in \mathcal{(NM)}\ \Leftrightarrow\ |h_{i},\ t\rangle_{F}\neq |h_{i},\ t\rangle_{I} \wedge i\in R
\end{equation}
According to this definition those sites that form the exterior
boundary of a relaxation process, whose states change without
toppling, are excluded from the corresponding $\mathcal{NM}$
structure.

\begin{figure}[!]
\scalebox{0.5}{\includegraphics[70pt,200pt][550pt,590pt]{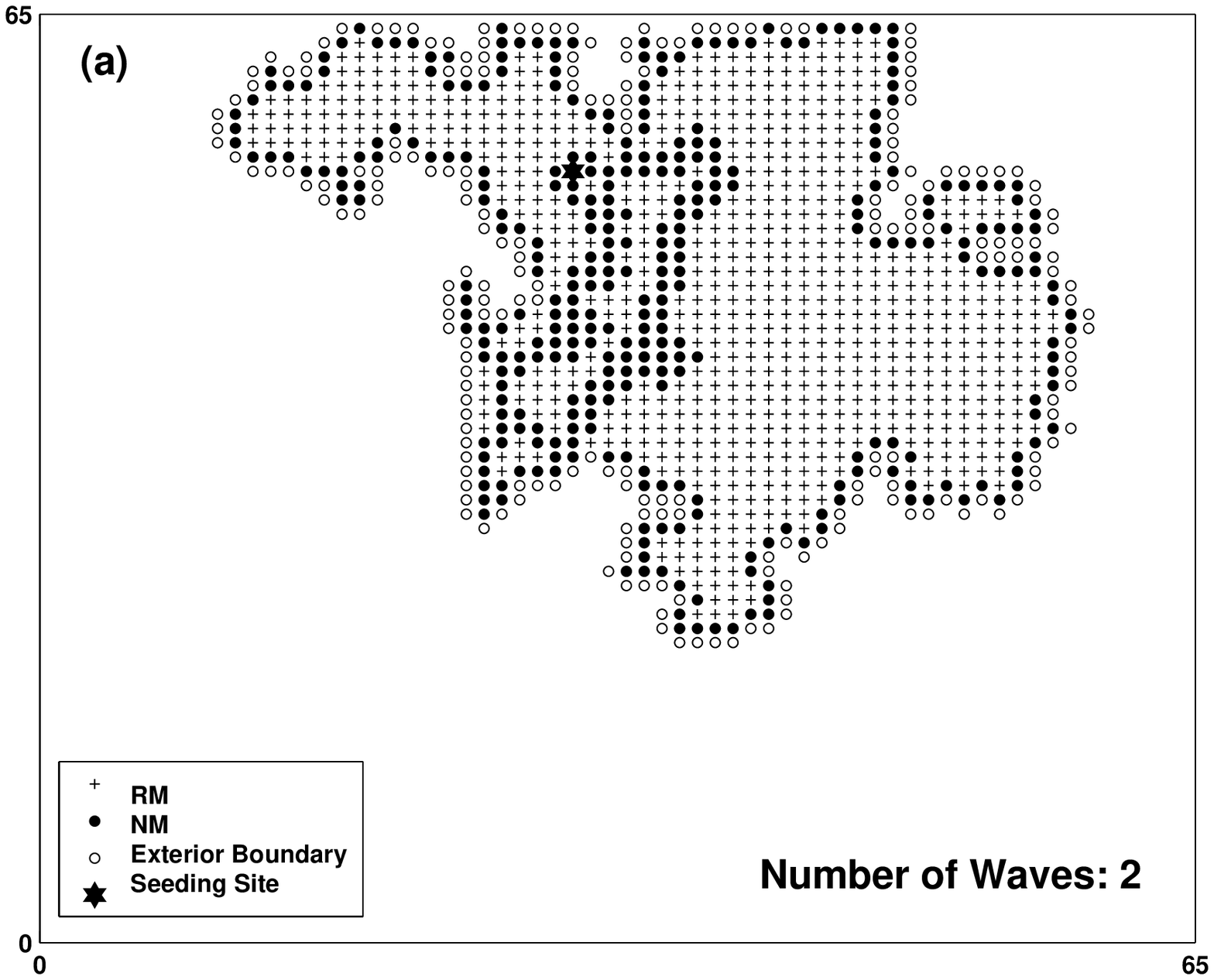}}
\scalebox{0.52}{\includegraphics[70pt,200pt][550pt,590pt]{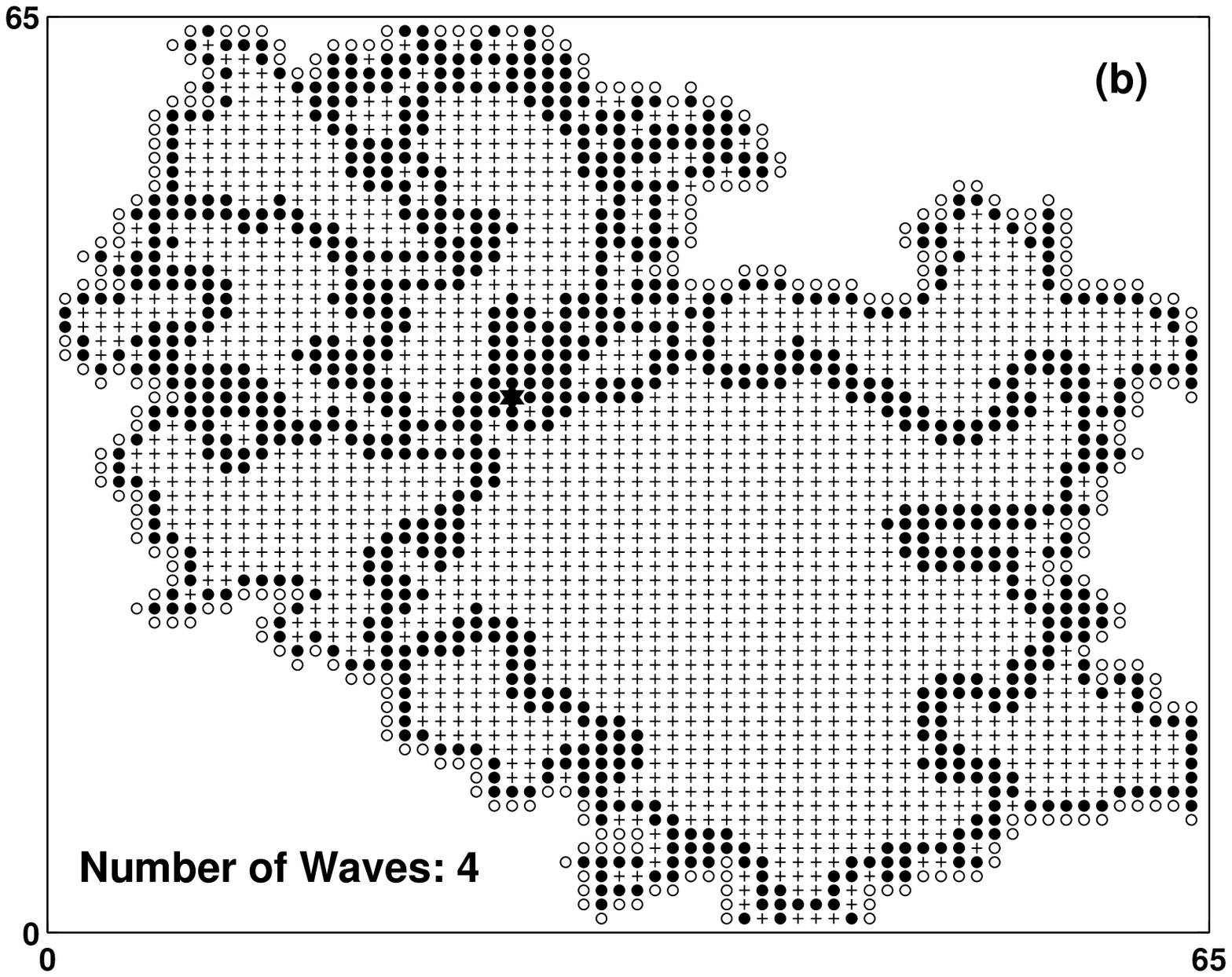}}
\caption{(a) Fine structure and exterior boundary of a simple
avalanche composed of two waves, and (b) a complex avalanche
composed of 4 waves in a $64 \times 64$ lattice. Here, the $(+)$,
$(\bullet)$ and $(\circ)$ signs represent $\mathcal{RM, \ NM,}$
and the exterior boundary of the avalanche, respectively. Note the
simple pattern of the $\mathcal{NM}$ structure in 1(a) compared to
that of 1(b).}
\end{figure}

Let $n_{\mathcal{RM(NM)}}$ be the size of a recurrent
(non-recurrent) macrostate, i.e. the total number of single site
recurrent (non-recurrent) microstates. Then for the area of the avalanche we have,
\begin{equation}a=n_{\mathcal{RM}}+n_{\mathcal{NM}}.\end{equation}
When the internal fine structure of an avalanche is made up of
multiple, spatially contained waves, the $\mathcal{NM}$
constitutes a thin layer on the boundary of each wave, while the
bulk of the relaxation process is mainly made up of a
$\mathcal{RM}$ structure. On the other hand, when an avalanche is
composed of multiple, interpenetrating waves, this simple
structure might be lost, and we may observe a complex, interweaved
pattern of $\mathcal{NM}$ and $\mathcal{RM}$ structures. These two
scenarios are shown in Figs. 1(a) and 1(b), respectively.

\section{Complexity in the fine structure of Avalanches}
In Ref.\ [29] it has been shown that the boundaries of consecutive
waves are not simply related to each other. In fact, the complexity
present in the fine structure of an avalanche is a direct
consequence of the ``complex superposition'' of the boundaries of
successive waves making up an avalanche. By ``complex
superposition'' we mean that in general in an avalanche,
excluding the exterior boundary,
\begin{equation}
    n_{\mathcal{NM}}\neq\sum_{k}(n_{\mathcal{NM}})_{\mathcal{W}_{k}}.
\end{equation}
Here and in the following we use symbols with subscript
$(\mathcal{W})$ to denote those quantities pertaining to a wave.
\textit{k} is the index of the waves making up a given avalanche.
Equation (7) simply states that the non-recurrent macrostate
($\mathcal{NM}$) of an avalanche is not the simple sum of the wave
boundaries which form the avalanche. So, the boundaries of
successive waves can interact and interpenetrate each other. Due
to this interaction, the fine structure of an avalanche may show a
complex pattern.

A quantity that contains valuable information about the dynamical
processes underlying the formation of the fine structure of a
relaxation process is the ratio of the size of $\mathcal{NM}$ to
$\mathcal{RM}$, i.e.\ $n_{\mathcal{NM}}/n_{\mathcal{RM}}$. In
Fig.\ 2, we have compared the conditional expectation value of
this quantity for avalanches and waves of a given area,
$E(n_{\mathcal{NM}}/n_{\mathcal{RM}}|a)$. While for avalanches of
intermediate size the ratio of $n_{\mathcal{NM}}/n_{\mathcal{RM}}$
tends to decrease, for larger ones we observe a gradual increase
in this quantity, indicating a crossover in the avalanches'
behavior, a point which we will return to later in this article.
As can be seen from the figure, the conditional expectation value
for the waves simply decreases with increasing area. This simply
shows that the boundary to bulk ratio for waves decreases with
increasing of the wave's size, as should be expected.
\begin{figure}[!]
\scalebox{0.5}{\includegraphics[45pt,195pt][540pt,585pt]{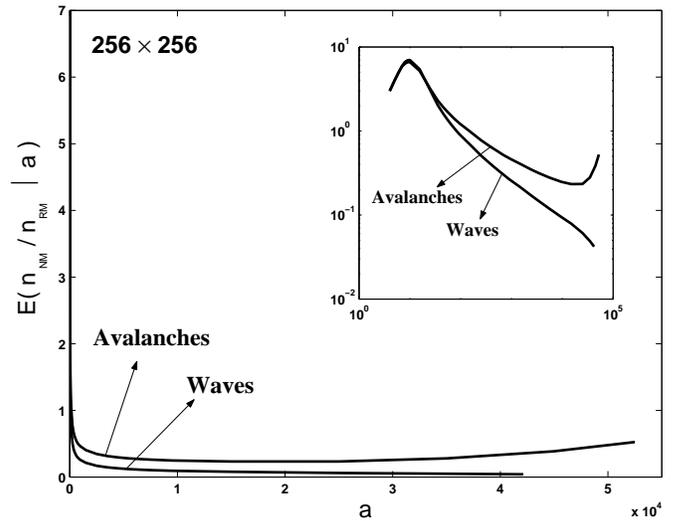}}
\caption{Comparison between the conditional expectation value of
$n_{\mathcal{NM}}/n_{\mathcal{RM}}$ in a system of linear size
L=256 for avalanches and waves of a given area. To reduce the
statistical fluctuation at large areas we have binned the data
logarithmically. This will result in losing a fraction of large
avalanches with
$E(n_{\mathcal{NM}}/n_{\mathcal{RM}}|a)\geqslant1$, which manifest
themselves when one considers the unbinned data. Inset shows the
same quantity on a double logarithmic graph. The bumps in the left
hand side of the inset are due to the grid properties of the
lattice.}
\end{figure}
However, in order to understand the nature of the observed
crossover phenomenon and the differences between the behavior of
waves of toppling and avalanches, we must study more fundamental
quantities describing dynamical properties of the critical state.
From Eq.\ (6) we can write the probability of having an avalanche
of area \textit{a} as,
\begin{equation}
P(a)=\sum_{n_{\mathcal{NM}},\ n_{\mathcal{RM}}}^{\ \ \prime} \! \!
\!
\!\!\!\!P(n_{\mathcal{RM}})P(n_{\mathcal{NM}}|n_{\mathcal{RM}}),
\end{equation} where the summation goes over those values
of $n_{\mathcal{NM(RM)}}$ that fulfill Eq.\ (6).  Here,
$P(n_{\mathcal{RM}})$ is the probability of having an avalanche
with a $\mathcal{RM}$ structure of size $n_{\mathcal{RM}}$.
$P(n_{\mathcal{NM}}|n_{\mathcal{RM}})$ is the conditional
probability distribution function (CPDF) of having a
$\mathcal{NM}$ structure of size $n_{\mathcal{NM}}$ for a
particular value of $n_{\mathcal{RM}}$. This equation, through the
quantity $P(n_{\mathcal{NM}}|n_{\mathcal{RM}})$, establishes a
direct connection between properties of the critical state and
dynamical aspects of the relaxation processes.  In Fig.\ 3, we
have plotted the CPDF,\ $P(n_{\mathcal{NM}}|n_{\mathcal{RM}})$ for
several values of $n_{\mathcal{RM}}$.\ We note that this function
is asymmetric about its maximum value. However, more importantly,
we observe the emergence of another local maximum with increasing
$n_{\mathcal{RM}}$ (point (b)).\ This suggests the emergence of a
different type of behavior as $n_{\mathcal{RM}}$ (or avalanche
size) increases.

What possible classification of avalanches can distinguish between
the two peaks in Fig.\ 3, and to what extent is it related to the
complexity of the fine structures of relaxation processes?\ If we
consider a similar quantity in the simple $\mathcal{NM}$, like the
$\mathcal{NM}$ of waves, our simulation shows that although the
asymmetric form of the CPDF persists, there is no significant
change in the general form of this function, i.e.\ no local
maximum emerges as $n_{\mathcal{RM}}$ is increased, see Fig.\ 4.
Therefore, in analogy with simple $(\mathcal{NM})$ structure of
waves, we define an avalanche to be simple, i.e. of simple fine
structure, if successive waves making up that avalanche are
spatially contained in each other, i.e.\ their $\mathcal{NM}$
structures (boundaries) are not interpenetrating, see Fig.\ 1(a).
In this case we have
$n_{\mathcal{NM}}\approx\sum_{k}(n_{\mathcal{NM}})_{\mathcal{W}_{k}}$.
Restrictly speaking, this definition does not exclude the
possibility of a weak interaction (mixing) between the
$\mathcal{NM}$ structures of a wave and that of its predecessor,
so that in general Eq.\ (7) holds. However, it is the extent of
the violation of the equality which categorizes avalanches into
simple or complex. In mathematical notation, for a sequence of
waves, $\mathcal{W}_{k},\ k=1,\ldots,n$, making up an avalanche we
have,\[Simple\ Fine\ Structure\ \Leftrightarrow \ \forall i \in
\mathcal{W}_{k}\Rightarrow i\in \mathcal{W}_{k-1}.\] On the other
hand, we consider a fine structure as complex, if the boundaries
of successive waves exceed the confines of their predecessors, see
Fig.\ 1(b).
\[Complex\ Fine\ Structure\ \Leftrightarrow \ \exists i \in
\mathcal{W}_{k} \ni i \not\in \mathcal{W}_{k-1}.\] We call these
two classes of avalanches type $\alpha$ and type $\beta$. As shown
in Fig.\ 3, our numerical simulation indicates that the CPDF
$P(n_{\mathcal{NM}}|n_{\mathcal{RM}})$\ is a superposition of
CPDF's of type-$\alpha$ and type-$\beta$ avalanches, in which the
first and second maxima correspond to type $\alpha$ (Fig.\ 3,
point (a)), and type-$\beta$ avalanches (Fig.\ 3, point (b)),
respectively.

\begin{figure}
\scalebox{0.5}{\includegraphics[100pt,185pt][535pt,600pt]{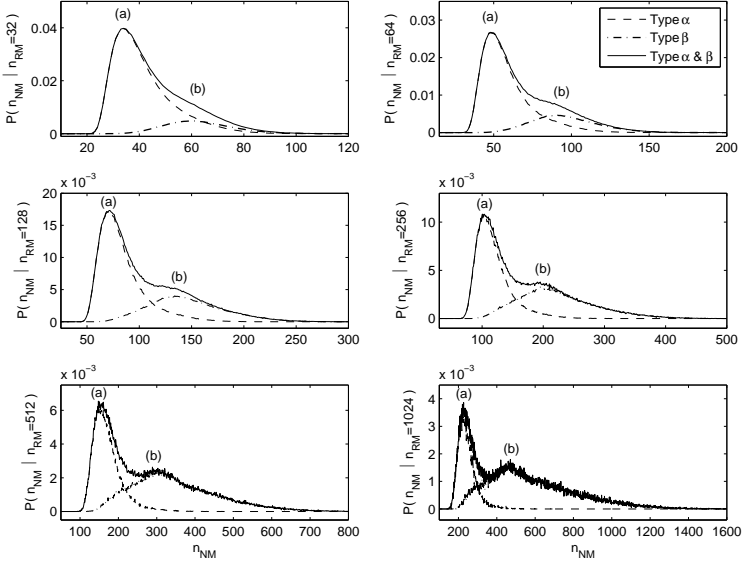}}
\caption{Normalized CPDF $P(n_{\mathcal{NM}}|n_{\mathcal{RM}})$ of avalanches as a function of $n_{\mathcal{NM}}$ for
different values of $n_{\mathcal{RM}}$ on a $512\times512$ lattice.
Note the emergence of a local maximum with increasing
$n_{\mathcal{RM}}$. This indicates the emergence of a different sort of
behavior with increasing avalanche size. Here the dashed and dash-dotted
lines show the decomposition of the CPDF to that of type $\alpha$ and type $\beta$ avalanches.}\end{figure}
\begin{figure}
\scalebox{0.5}{\includegraphics[100pt,190pt][530pt,610pt]{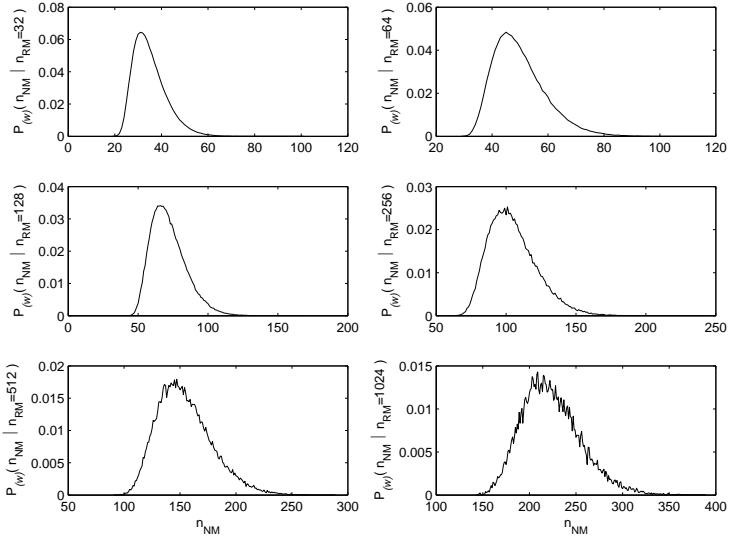}}
\caption{Normalized CPDF
$P_\mathcal{(W)}(n_{\mathcal{NM}}|n_{\mathcal{RM}})$ of waves as a
function of $n_{\mathcal{NM}}$ for different values of
$n_{\mathcal{RM}}$. No local maximum emerges as $n_{\mathcal{RM}}$
is increased.}
\end{figure}

At this point, we need to make a connection between the properties
of the critical state and those of the type $\alpha$ and type
$\beta$ avalanches. Let \textit{N} be the total number of
avalanches. Then, \mbox{$N=N_{\alpha}+N_{\beta}$} where
$N_{\alpha}$ and $N_{\beta}$ are the number of type $\alpha$ and
type $\beta$ avalanches, respectively. Using the abbreviated
notations $n_{\mathcal{NM}}\mapsto y$ and $n_{\mathcal{RM}}\mapsto
x$, we have the following definitions:
\begin{itemize}
   \item $N^{\alpha(\beta)}$: Total number of type $\alpha$
   (or $\beta$) avalanches.
   \item $N_{yx}^{\alpha(\beta)}$: Total number of type $\alpha$
   (or $\beta$) avalanches with $\mathcal{NM}$ of size $y$ and
   $\mathcal{RM}$ of size $x$.
   \item $N_{y}^{\alpha(\beta)}$: Total number of type $\alpha$
   (or $\beta$) avalanches with $\mathcal{NM}$ of size $y$.
   \item $N_{x}^{\alpha(\beta)}$: Total number of type $\alpha$
   (or $\beta$) avalanches with $\mathcal{RM}$ of size $x$.
   \item $N_{x}$: Total number of avalanches with $\mathcal{RM}$ of size $x$.
   \item $N_{yx}$: Total number of avalanches with $\mathcal{NM}$ of size $y$ and $\mathcal{RM}$ of size $x$.
\end{itemize}
Using these definitions, we can rewrite Eq.\ (8) in terms of the
properties of type $\alpha$ and type $\beta$ avalanches.  We start
with:
\begin{eqnarray}
    P(y|x)&=&\frac{N_{yx}}{N_{x}}=\frac{N_{yx}^{\alpha}+N_{yx}^{\beta}}{N_{x}}=\frac{N_{yx}^{\alpha}}{N_{x}^{\alpha}}\frac{N_{x}^{\alpha}}{N_{x}}+\frac{N_{yx}^{\beta}}{N_{x}^{\beta}}\frac{N_{x}^{\beta}}{N_{x}} \nonumber \\
    &=&P_{\alpha}(y|x)\frac{N_{x}^{\alpha}}{N_{x}}+P_{\beta}(y|x)\frac{N_{x}^{\beta}}{N_{x}}.
\end{eqnarray}
Substituting Eq.\ (9) in (8), and using the fact that
$P(x)=N_{x}/N$ we have,
\begin{equation}
P(a)=\sum_{x,\ y}^{\ \ \prime}P(x)P(y|x)=\sum_{x,\ y}^{\ \
\prime}\left[P_{\alpha}(y|x)\frac{N^{\alpha}_{x}}{N}+P_{\beta}(y|x)\frac{N^{\beta}_{x}}{N}
\right],
\end{equation}
where the summation goes over those values of $x$ and $y$ that
fulfill Eq.\ (6). Finally, using the relation,
\[N_{x}^{\alpha(\beta)}/N=\frac{N_{x}^{\alpha(\beta)}}{N^{\alpha(\beta)}}\frac{N^{\alpha(\beta)}}
{N}=P_{\alpha(\beta)}P_{\alpha(\beta)}(x),\]
we have,
\begin{equation}
P(a)=\sum_{x,\ y}^{\ \
\prime}[P_{\alpha}P_{\alpha}(x)P_{\alpha}(y|x)+P_{\beta}P_{\beta}(x)P_{\beta}(y|x)].
\end{equation}

According to Eq.\ (11), an avalanche size distribution function
can be written as a separate combination  of type-$\alpha$ and
type-$\beta$ distribution functions. Now, if type-$\alpha$ and
type-$\beta$ avalanches have similar scaling properties, one can
expect the avalanche-size probability distribution functions to
scale accordingly. However, if these two types of avalanches have
different and distinct scaling properties, one cannot find a
consistent scaling behavior for the overall probability
distribution function. This is in fact the key message of the
present Article. In the next section we will give analytical as
well as numerical evidence on how the scaling properties of these
two types of avalanches differ from each other.  The key
difference, as we will see, is the size-dependence of the scaling
behavior in type-$\beta$ avalanches.

\section{Scaling of CPDF For Different Types of Avalanches}
In order to investigate the properties of $P(a)$ it is important
to study the properties of $P_{\alpha}(y|x)$ and $P_{\beta}(y|x)$.
We have carried out an extensive study of such CPDF's. We find
that the scaling behavior of these distributions are distinctly
different. Accordingly, we propose the following scaling
ans\"{a}tz for the above distribution functions:
\begin{itemize}
\item Type $\alpha$ avalanches:
\begin{equation}
P_{\alpha}(y|x)=x^{-\gamma_{\alpha}}U_{\alpha}(\frac{y-E_{\alpha}(y|x)}{x^{\gamma_{\alpha}}}).
\end{equation}
\item Type $\beta$ avalanches:
\begin{eqnarray}
P_{\beta}(y|x)&=&x^{-\gamma_{\beta}}U_{\beta}(\frac{y+x+E_{\beta}(y|x)}{x^{\gamma_{\beta}}}),
\end{eqnarray}
\end{itemize}
where $\gamma_\beta$ is a size-dependent exponent, i.e. $\gamma_\beta=\gamma_\beta(x)$.
Here $U_{\alpha}$ and $U_{\beta}$ are universal functions, and
$E_{\alpha(\beta)}(y|x)$ is the mean value of the given CPDF,
defined through the relation $E_{\alpha(\beta)}(y|x)=\int
yP_{\alpha(\beta)}(y|x)\,dy$.

From Eqs.\ (12) and (13), we can readily calculate the scaling
behavior of the first and second moments of $y$. This provides a
suitable way via which one can confirm the proposed scaling
ans\"{a}tz for the corresponding CPDF's.

Let us first consider the case of type $\beta$ avalanches. Using
the suggested form in Eq.\ (13), we have
\begin{eqnarray}
E_{\beta}(y|x)&=&\int_{0}^{\infty}yP_{\beta}(y|x)\ dy \\
&=&x^{-\gamma_{\beta}}\int_{0}^{\infty}yU_{\beta}((y+x+E_{\beta}(y|x))/x^{\gamma_{\beta}})\
dy. \nonumber
\end{eqnarray}
Performing the change of variable $z=y+x+E_{\beta}(y|x)$ in Eq.\
(14) and integrating we obtain,
\begin{equation}
E_{\beta}(y|x)=\frac{1}{2}Cx^{\gamma_{\beta}}-\frac{x}{2},
\end{equation}
where $C=\int \xi\,U_{\beta}(\xi)\ d\xi;\
\xi=z\,x^{-\gamma_{\beta}}$. So, after the addition of the linear
term $x/2$ to $E_{\beta}(y|x)$, it must scale as
$x^{\gamma_\beta(x)}$ for different scaling regions defined by the
area or the size of the $\mathcal{RM}$ structure of an avalanche.
In the case of type $\alpha$ avalanches, we cannot obtain the
scaling behavior of the corresponding conditional expectation
value, $E_{\alpha}(y|x)$ from $P_{\alpha}(y|x)$, as we did in the
case of type $\beta$ avalanches. So, to obtain any further
information, we must look at higher moments of $y$, e.g.\
$E_{\alpha}(y^{2}|x)$.
\begin{eqnarray}
E_{\alpha}(y^{2}|x)&=&\int_{0}^{\infty}\!\!\!y^{2}P_{\alpha}(y|x)\,dy \\
&=&x^{-\gamma_{\alpha}}\int_{0}^{\infty}\!\!\!y^{2}\,U_{\alpha}((y-E_{\alpha}(y|x))/x^{\gamma_{\alpha}})\,dy.
\nonumber
\end{eqnarray}
Performing the change of variable $z=y-E_{\alpha}(y|x)$, we can
rewrite Eq.\ (16) as,
\begin{equation}
E_{\alpha}(y^{2}|x)=C\,x^{2\,\gamma_{\alpha}}+[E_{\alpha}(y|x)]^{2},
\end{equation}
where $C=\int \xi^{2}\,U_{\alpha}(\xi)\ d\xi;\
\xi=z\,x^{-\gamma_{\alpha}}$. Therefore,
\begin{equation}
\sigma_{\alpha}=\left[E_{\alpha}(y^{2}|x)-(E_{\alpha}(y|x))^{2}\right]^{1/2}\sim
x^{\gamma_{\alpha}},
\end{equation}
where $\sigma_{\alpha}$ is the standard deviation of the given
distribution. In obtaining the last relation we have used the fact
that $\int \xi U_{\alpha}(\xi)\,d\xi=0$. So, in the case of type
$\alpha$ avalanches $\sigma_{\alpha}$ scales as
$x^{\gamma_{\alpha}}$.

\begin{figure}
\scalebox{0.45}{\includegraphics[80pt,180pt][520pt,585pt]{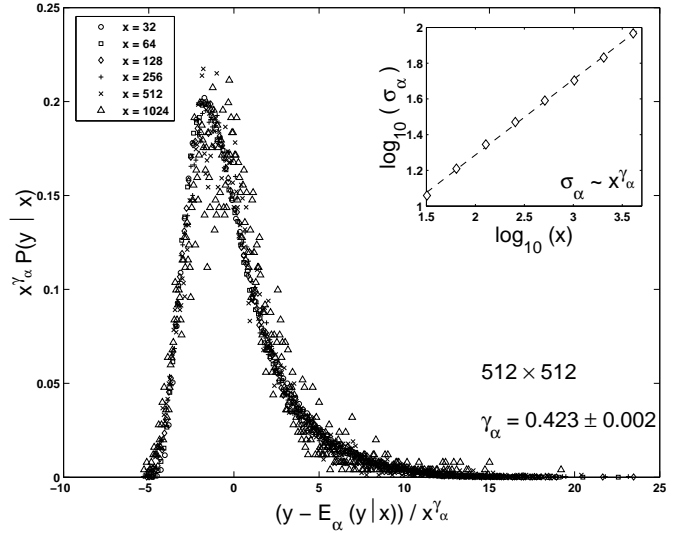}}
\caption{CPDF data collapse using the suggested form, Eq.\ (12),
for type-$\alpha$ avalanches. Here, we obtain a good collapse for
different values of $n_{\mathcal{RM}}$ with
$\gamma_{\alpha}=0.423\pm0.002$. Inset shows the scaling of
$\sigma_\alpha$ with $x$ for different values of $x$ ranging from
$x=32$ to $x=4096$. The slope of the dashed line is equal to
$\gamma_\alpha=0.423$}
\end{figure}

\begin{figure}
\scalebox{0.46}{\includegraphics[100pt,130pt][550pt,590pt]{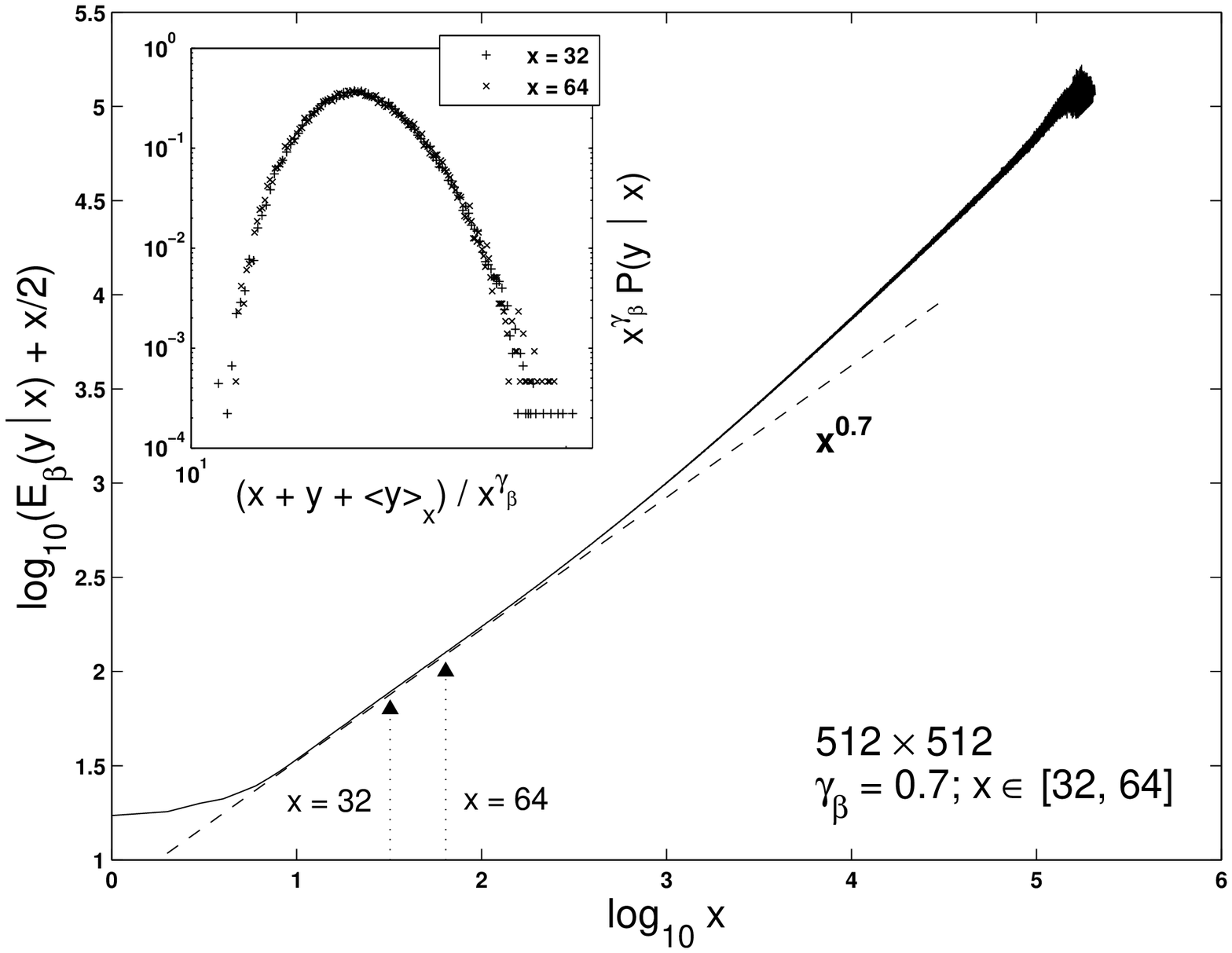}}
\scalebox{0.46}{\includegraphics[100pt,120pt][550pt,600pt]{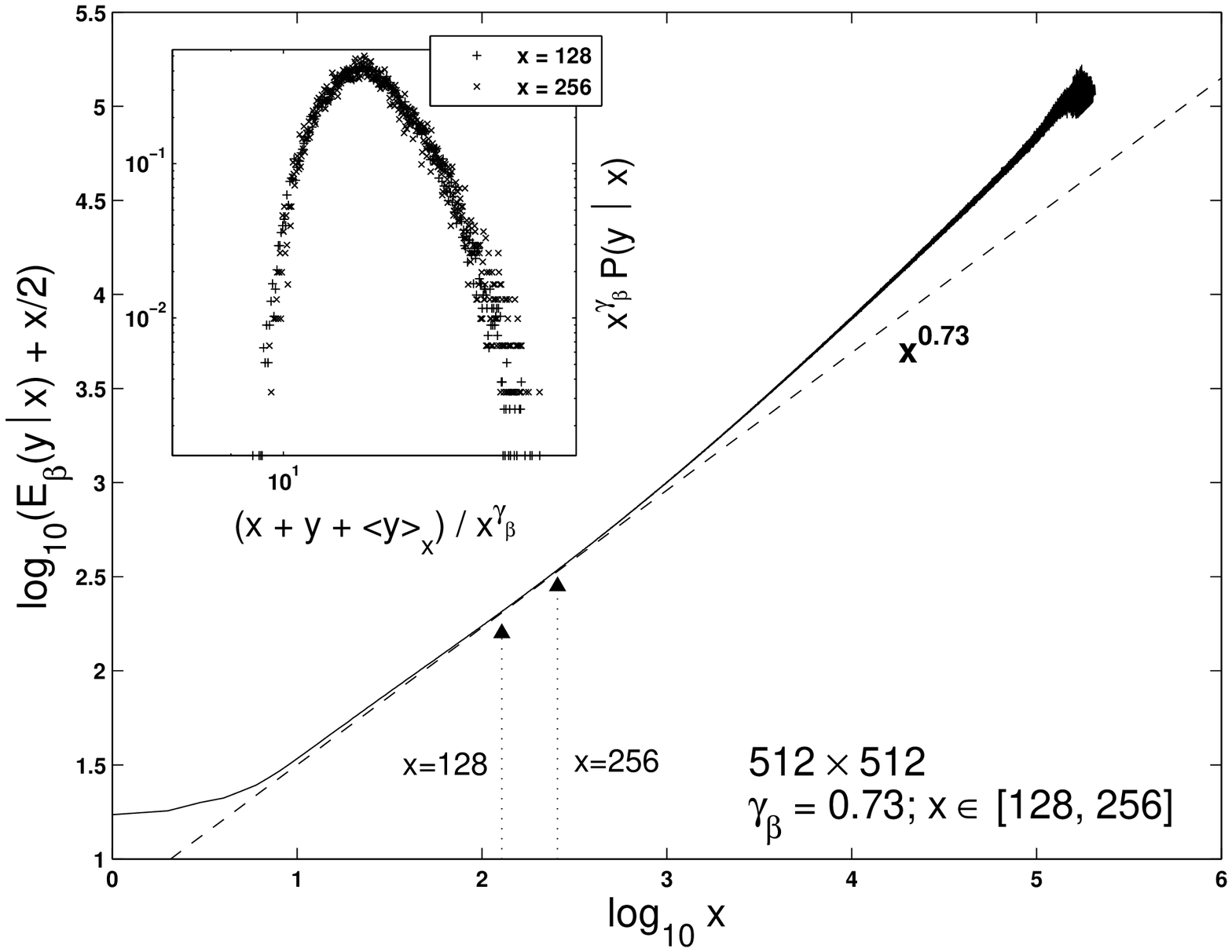}}
\scalebox{0.46}{\includegraphics[100pt,180pt][550pt,590pt]{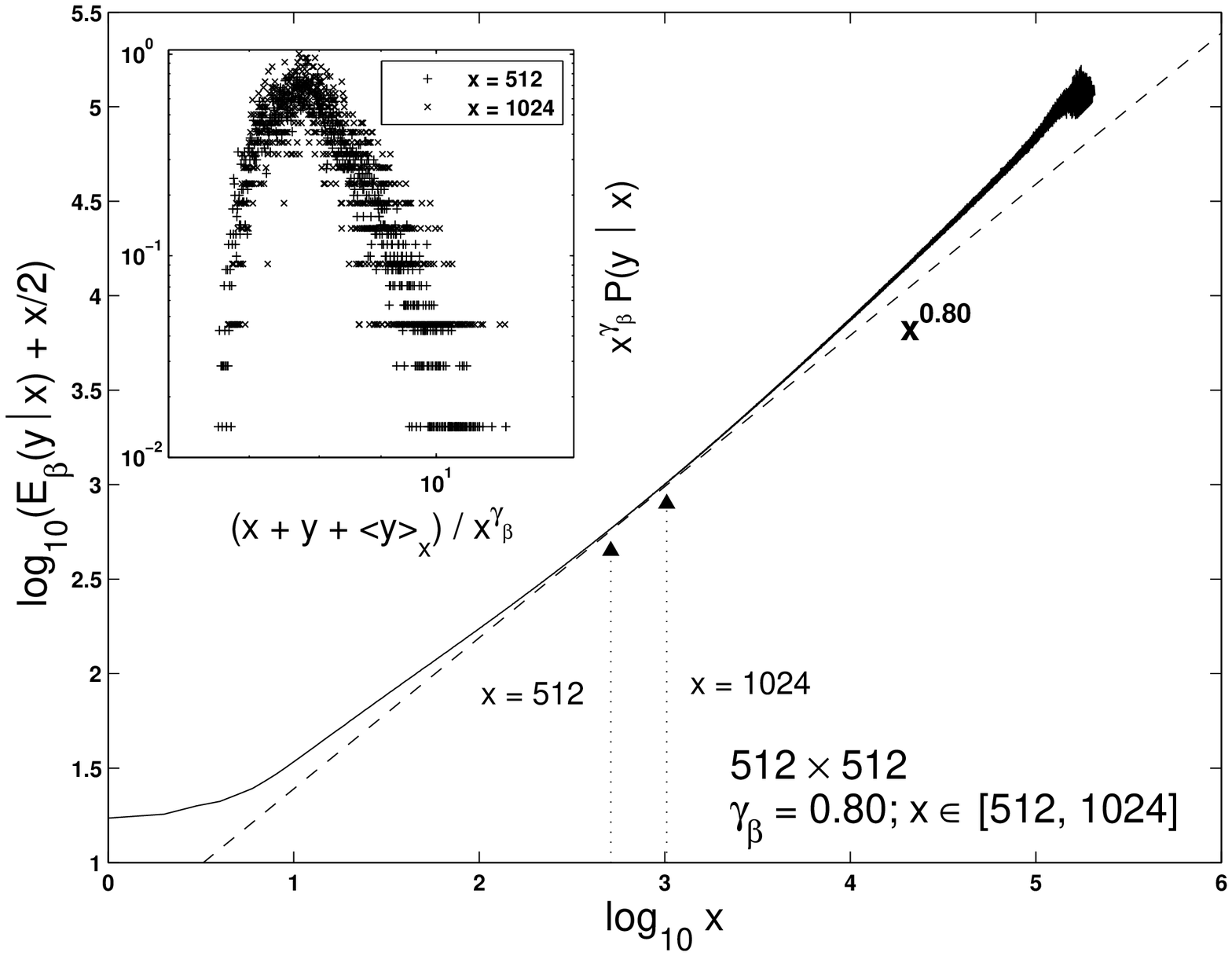}}
\caption{Scaling behavior of the conditional expectation value, $E_{\beta}(y|x)$,
after the addition of the linear term $x/2$ in a system of linear
size $L=512$ for (a) small avalanches $(32\leq x \leq64)$,(b)
intermediate avalanches $(128\leq x \leq 256)$, (c) and large
avalanches $(512\leq x \leq 1024)$. Insets show the collapse of
CPDF, $P_{\beta}(y|x)$, using the suggested form, Eq.\ (13), for
different scaling regions, which determines the value of the
scaling exponent $\gamma_{\beta}(x)$. Note how $\gamma_{\beta}$
increases with avalanche size.}
\end{figure}

To verify the proposed scaling forms and also to extract the
scaling exponents $\gamma_{\alpha}$ and $\gamma_{\beta}$, we
implement the method of data collapse in line with Ref.\ [25]. In
Fig.\ 5, we have applied this technique to type $\alpha$
avalanches for different values of $x(=n_{\mathcal{RM}})$. As can
be seen, we obtain a reasonable collapse with the scaling exponent
$\gamma_{\alpha}=0.423$. The inset shows the scaling of standard
deviation, $\sigma_\alpha$, with $x$ for the corresponding CPDF's.
There the dashed line shows $x^{\gamma_\alpha}$. This verifies in
a straightforward manner, the validity of Eq.\ (18), thus lending
support to our scaling ans\"{a}tz, Eq.(12). By performing a
similar analysis on an ensemble of waves, we could verify that the
CPDF $P_{\mathcal{(W)}}(y \mid x)$ of waves of toppling also
possesses a similar scaling to type $\alpha$ avalanches, Eq. (12),
with the scaling exponent $\gamma_{\mathcal{(W)}}=0.47$. This
indicates an interesting similarity between the scaling behavior
of these two types of relaxation events. However, the obvious
difference in the values of the scaling exponents,
$\gamma_{\alpha}$ and $\gamma_{(\mathcal{W})}$, reveals the
essential differences in the nature of these two relaxation
processes. In other words, one cannot consider type-$\alpha$
avalanches as a simple sum of waves making up that avalanche.

Now, the situation is not as simple for type-$\beta$ avalanches.
We observe that we cannot find a unique exponent $\gamma_{\beta}$,
which collapses our data for all values of $x(=n_{\mathcal{RM}})$.
Instead, we find that $E_{\beta}(y|x)+x/2$ does not have a unique
slope (on a logarithmic plot) and shows an obvious curvature as
can be seen in Fig.\ 6.  Here instead, we find different scaling
exponents for different scaling regions.  We divide our scaling
region into small, intermediate, and large avalanches and perform
our collapse (the insets in Fig.\ 6) with different exponent for
each of these regions. This is shown in different parts of Fig.\
6.  The important and fundamental difference here in the case of
$\beta$-avalanches is that $\gamma_{\beta}$ is size-dependent and
increases with avalanche's size. However, this increase cannot be
unbound and $\gamma_\beta$ should eventually saturate. Our
numerical results show that this exponents saturates at
$\gamma_\beta=1$, indicating a range of $0.7\leq
\gamma_{\beta}\leq 1.0$ for this exponent, see Fig.\ 7. Since we
do not expect that the boundary of a relaxation process grows
larger than its bulk, the value $\gamma_\beta(x>>1)=1$ is in fact
the physical upper limit for this exponent. Another way to see
that $\gamma_{\beta}$ saturates at $\gamma_{\beta}=1.0$ is to plot
$E_{\beta}(\frac{y}{x}|x)+1/2$, which according to our scaling
ans\"{a}tz should scale as $x^{\gamma_{\beta}(x)-1}$.  This is
shown as an inset in Fig.\ 7. One can see this saturation as the
eventual $x$-independence of the plot for large avalanches.  We
therefore conclude that $\gamma_{\beta}$, unlike
$\gamma_{\alpha}$, does not have a fixed value and in fact varies
between $0.7$ and $1.0$ with increasing avalanche size.  Such form
of avalanche-size dependent exponent is a result of the complexity
inherent in the dynamical behavior of type-$\beta$ avalanches.

It is now clear why simple FSS fails in the 2-D BTW ASM.
Avalanches can be categorized in two classes each of which has
distinctly different scaling properties.  The combination of these
two cannot exhibit a consistent scaling behavior.  Moreover, the
culprit is identified as type-$\beta$ avalanches for which the
scaling exponent depends on the avalanche size. The coexistence of
two different types of avalanches in the critical state, with
distinctly different scaling behavior, and the failure of FSS
picture, has been already observed and analytically proved in 1-D
ASM [32]. However, and to the best of our knowledge, this is for
the first time that this phenomenon is reported and numerically
verified in the 2-D BTW ASM. Note that this classification is
irrespective of the boundary and system size effects and is
inherent in the dynamical properties of an avalanche.

\begin{figure}[!]
\scalebox{0.55}{\includegraphics[115pt,180pt][545pt,600pt]{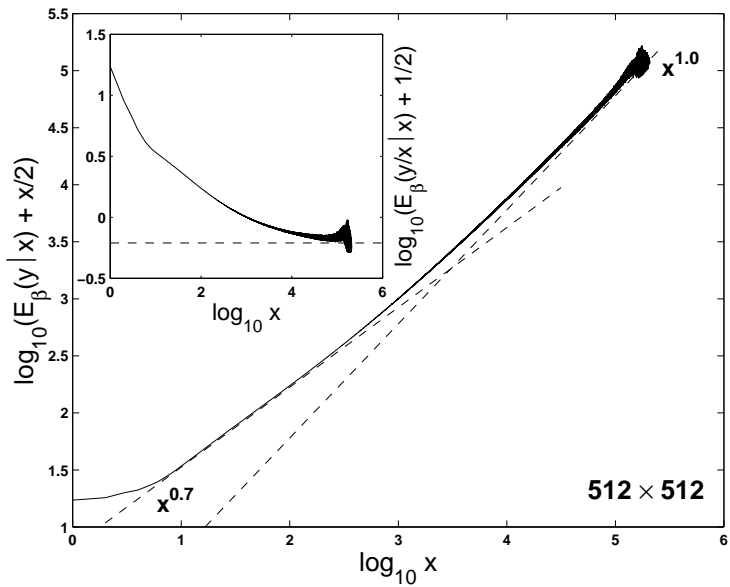}}
\caption{Conditional expectation value, $E_{\beta}(y|x)$, after the
addition of the linear term $x/2$ for a system of linear size
$L=512$. Note the curvature in the graph for different values of
$x$. Here the dashed lines have slopes 0.7 and 1.0 respectively.
Inset shows the scaling behavior of $E_{\beta}(\frac{y}{x}|x)+1/2$
for different values of $x$. The gradual increase of
$\gamma_{\beta}(x)$ to the value $\gamma_{\beta}(x)=1$ is clear. The
horizontal dashed line is as a reference for sight.}
\end{figure}
\begin{figure}[!]
\scalebox{0.5}{\includegraphics[115pt,180pt][545pt,600pt]{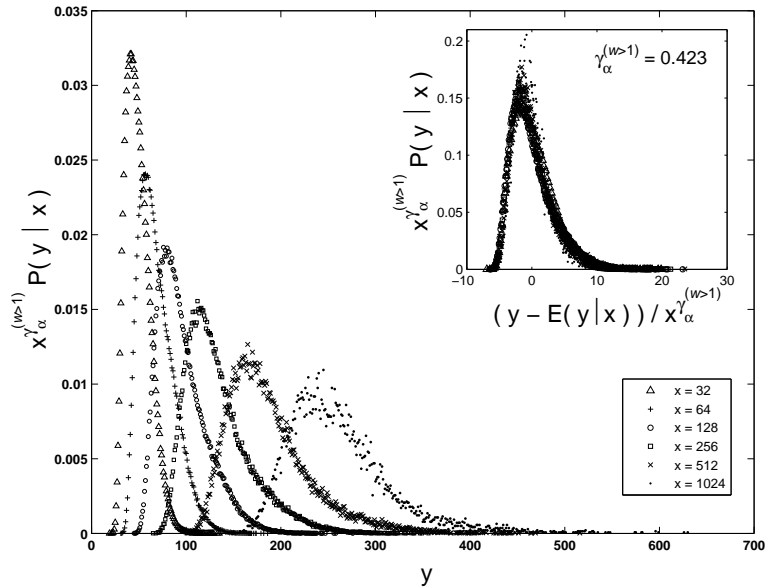}}
\caption{CPDF of type $\alpha$ avalanches with more than one wave
for different values of $n_{\mathcal{RM}}$ in a $512\times512$
lattice. Inset shows the collapse of CPDF of these avalanches using
the suggested form of the CPDF of type $\alpha$ avalanches. Here
again one obtains a reasonable collapse with
$\gamma_{\alpha}^{(\mathit{\mathcal{W}}>1)}=0.423$.}
\end{figure}

Here, one might wonder if our distinction of type-$\alpha$ and
-$\beta$ avalanches is merely an analysis of avalanches made up of
a simple wave versus a collection of waves. In order to address
this issue we have performed the same analysis on the properties
of type-$\alpha$ avalanches made up of more than one wave. Our
numerical analysis shows that, despite the existence of multiple
toppling events (waves) in this class of avalanches, there is no
dependency in their scaling behavior on the $\mathcal{NM}$
structure (or area) of the avalanches. In fact, as it is shown in
Fig.\ 8, this type of avalanches collapse with the same exponent
as that of type-$\alpha$ avalanches shown in Fig.\ 5. We now can
assert that what distinguishes $\alpha$- and $\beta$-avalanches is
the interaction (mixing) of wave boundaries and not the mere
existence of multiple waves. In other words, simple waves make up
avalanches of type-$\alpha$ and complex (mixing) waves constitute
complex or $\beta$-avalanches.

Finally, of particular importance is the relation between the
behavior of the scaling exponents $\gamma_{\alpha(\beta)}$ and the
deviation from simple power law behavior in the quantities such as
$P(s)$, where $s$ is the size (or total number of topplings) in an
avalanche, in the 2-D BTW model. In order to gain a better
understanding of this general behavior, we have also looked at
some other SOC models in this regard. Our preliminary results in
the case of Manna model, which obeys FSS, show that the scaling
exponent $\gamma$ maintains a constant value close to 0.5 [31, 33,
34]. Such observation, as well as the similarities between the
behavior of waves and type-$\alpha$ avalanches, makes us believe
that the constancy of the $\gamma$ exponent indicates that the
scaling behavior of the corresponding relaxation process follows a
simple power law behavior. In the case of size dependency of the
exponent $\gamma$, it is the rate of change of $\gamma$ which
determines whether the power law behavior emerges or not. For
example, in the particular case of the type-$\beta$ avalanches in
the present model, as $\gamma_\beta$ approaches unity for large
avalanches, the rate of change of $\gamma_\beta$ becomes very
small, as can be seen from Figs.\ 6 and 7. Therefore, we expect
that the power law behavior is recovered for large type-$\beta$
avalanches. While this reasoning holds true for quantities like
$P(s)$ in which multiple toppling events play an important role,
one should take much more care in interpreting the behavior of
other quantities such as $P(a)$. In fact, due to the large
fluctuations in the size of a $\mathcal{NM}$ structure of a
$\beta$-avalanche (see Fig.\ 3), the delicate size dependency of
the exponent $\gamma_\beta$ is effectively blurred by performing
the summation in Eq.\ (11). In such situation, the effect of
averaging, as well as ignoring the fundamental differences between
the two types of avalanches, will result in a seemingly power law
behavior in the case of $P(a)$ [23]. Indeed, by gathering separate
statistics for $\alpha$- and $\beta$-avalanches, one can clearly
observe how the scaling behavior of quantities such as
$P_{\beta}(a)$ and $P_{\beta}(s)$ deviates from a simple power law
for small and intermediate avalanche sizes. However, the power law
behavior is recovered for large $\beta$-avalanches [31].

\section{Conclusion}
To summarize, in this Article, we have shown that avalanches in
the ASM can have complex fine structures as a result of
interaction of wave boundaries within a given avalanche.  We used
these interactions to define simple (type-$\alpha$) and complex
(type-$\beta$) avalanches.  We have studied the scaling behavior
of these two avalanche types in detail and have highlighted their
differences. We have shown how one can view the general dynamical
scaling properties of this model in terms of scaling properties of
the combined type-$\alpha$ and type-$\beta$ avalanches.  We have
proposed scaling ans\"{a}tz for these two types of avalanches and
have verified them numerically, thus showing how these two types
of avalanches have distinctly different scaling behavior. In
particular, while type-$\alpha$ avalanches are characterized by a
constant-value scaling exponent, type-$\beta$ avalanches are
characterized by an avalanche-size dependent exponent.  We
believe, this distinction between type-$\alpha$ and type-$\beta$
avalanches underlies the failure of consistent (finite-size)
scaling in this model. Moreover, we argued how the size dependency
of the scaling exponent in $\beta$-avalanches leads to the
deviation from power law behavior in important quantities
describing the dynamical behavior of the avalanches, such as
$P(s)$. In addition, our results indicate that due to the
coexistence of two distinctly different relaxation events in the
critical state of the BTW model, one must separate these two
events in gathering any reliable statistics from the system. We
hope that our analysis is helpful in answering some of the long
standing problems on the behavior of the prototype model of SOC in
two dimensions. Moreover, we note that our classification of
avalanches opens up many questions as well.  For example, how are
the dynamical behavior of these avalanches different from each
other?  In particular, what are the essential characteristics of
wave boundary interactions?  Can such classifications be useful in
other models of SOC?  We plan to address some of these issues in a
forthcoming publication [31].
\section{Acknowledgement}
The authors would like to thank M.M. Golshan for many useful
conversations. Partial financial support of Shiraz University
Research Counsel is kindly acknowledged.

\end{document}